\documentclass[twocolumn,preprintnumbers,amsmath,amssymb,prl]{revtex4}

\usepackage{graphicx}
\usepackage{dcolumn}

\usepackage{mathptmx, courier, pifont}
\usepackage[scaled=0.92]{helvet}
\usepackage[T1]{fontenc}
\usepackage{textcomp}

\usepackage{bm}

\usepackage{color}

\newcommand{\singlefig}[6]{%
\begin{figure}\vspace{#3}%
\includegraphics*[scale=#5]{#2}%
\caption{\label{fig:#1} #6}%
\vspace{#4}%
\end{figure}}

\newtheorem{conjecture}{Conjecture}

\newcommand{\tsub}[1]{_{\mbox{\scriptsize#1}}}

\newcommand{\fig}[1]{Fig.~\ref{fig:#1}}

\newcommand{\isotope}[2]{\mbox{$^{#1}$#2}}

\begin{document}

\title{Superconductivity and Superfluidity as Universal Emergent Phenomena}

\author{Mike Guidry$^{(1)}$}
\email{guidry@utk.edu}
\author{Yang Sun$^{(2,3,4)}$}
\email{sunyang@sjtu.edu.cn}

\affiliation{
$^{(1)}$Department of Physics and Astronomy, University of
Tennessee, Knoxville, Tennessee 37996, USA
\\
$^{(2)}$Department of Physics and Astronomy, Shanghai Jiao Tong
 University, Shanghai 200240, People's Republic of China
\\
$^{(3)}$IFSA Collaborative Innovation Center, Shanghai Jiao Tong
University, Shanghai 200240, China
\\
$^{(4)}$State Key Laboratory of Theoretical Physics, Institute of
Theoretical Physics, Chinese Academy of Sciences, Beijing 100190,
China }

\date{\today}


\begin{abstract}
Superconductivity (SC) or superfluidity (SF)  is observed across a
remarkably broad range of fermionic systems: in BCS, cuprate,
iron-based, organic, and heavy-fermion superconductors, and
superfluid helium-3 in condensed matter; in a variety of SC/SF
phenomena in low-energy nuclear physics; in ultracold, trapped
atomic gases; and in various exotic possibilities in neutron stars.
The range of physical conditions and differences in  microscopic
physics  defy all attempts to unify this behavior in any
conventional picture. Here we propose a unification through the
shared symmetry properties of the emergent condensed states, with
microscopic differences absorbed into parameters. This, in turn,
forces a rethinking of specific occurrences of SC/SF such as cuprate
high-$T\tsub c$ superconductivity, which becomes far less mysterious
when seen as part of a continuum of behavior shared by a variety of
other systems.
\end{abstract}

\pacs{71.10.-w, 71.27.+a, 74.72.-h}

\maketitle

Superconductivity and superfluidity are collective phenomena owing
their existence to many-body interactions; the corresponding {\em
emergent states} are not related perturbatively to the parent state.
Thus, characterization of SC and SF through microscopic properties
of the parent system fails on two levels: (1)~It cannot provide a
unified view, since  microscopic physics differs fundamentally
between fields. (2)~The transition from the microscopic parent state
to the collective emergent state is not analytic; thus it is
conjecture to assume that microscopic tendencies of the parent state
are related directly to collective properties of the emergent state.

Conventional understanding of SC and SF is built on the idea of a
{\em Fermi liquid,} for which single-particle states of the
interacting system are  in one-to-one correspondence with those of
the non-interacting system. Superconductivity is assumed to develop
from a Fermi-liquid parent through the {\em Cooper instability,} in
which two fermions outside a filled Fermi sea can form a bound state
for vanishingly small attraction \cite{coop56}. In the solid state
the weak attraction is assumed conventionally to arise from
interaction of electrons with lattice vibrations.

The Cooper instability  was developed into a many-body theory by the
Bardeen--Cooper--Schrieffer (BCS) postulate that the SC state is a
coherent superposition of fermion pairs  in a weak coupling limit
\cite{BCS57}, and this was generalized to Eliashberg theory, which
removed the weak-coupling restrictions.  The BCS idea was soon
adapted to applications in nuclear physics \cite{bmp58}, with pairs
bound by attractive nucleon--nucleon forces.

BCS theory in condensed matter and  nuclear  physics involves quite
different interactions operating on energy and distance scales
differing by many orders of magnitude.  However, emergent SC/SF
properties were unified through sharing the same form for the BCS
wavefunction, which implied a common pseudospin symmetry of the
effective Hamiltonians that could be  expressed elegantly in terms
of an SU(2) Lie algebra \cite{ande58}. Thus similarities between
these fields could be understood through a common algebraic
structure, while differences could be viewed as primarily {\em
parametric} and not fundamental.

By the early 1970s all cases of fermionic SC and SF were thought to
be understood in these conventional BCS terms. This unity was
shattered by a series of discoveries beginning with  \isotope{3}{He}
superfluidity in 1972 \cite{oshe72}, followed by SC in heavy-fermion
compounds in 1979 \cite{steg79}, SC in various organics beginning in
1980 \cite{jero2012}, SC for  copper oxides at high temperatures in
1986 \cite{highTc_discovery}, high-temperature SC for various
iron-based compounds beginning in 2008 \cite{FeAsDiscovery} and, in
the past decade, direct observations suggesting proton SC and
neutron SF in neutron stars \cite{nstar}, and superfluidity for
ultracold fermionic atoms \cite{ultracold}.

It is thought that SC or SF in all these systems results from
condensation of Cooper pairs in parent states that may not be Fermi
liquids, through interactions that may not be mediated by phonons and
may differ from the  $s$-wave form of conventional BCS theory ({\em
unconventional pairing}). This calls into question whether the  BCS
paradigm, even generalized to accommodate unconventional pairing,
can describe the diversity of SC and SF behavior. The issue is how
the Cooper instability emerges from a variety of parent states that
need not be Fermi liquids, enabled by highly diverse
fermion--fermion correlations.

Let us begin with a brief survey of SC and SF behavior. Our aim is
to highlight the simultaneous microscopic diversity but
emergent-level unity of superconductivity and superfluidity. Tests
of conventional BCS are well known so we shall emphasize more
complex behavior, with  BCS viewed as a limit of this more complex
behavior.

Phase diagrams for cuprate superconductors  are rather universal,
with features
similar to those of \fig{phaseComposite}(a).%
\singlefig
{phaseComposite}       
{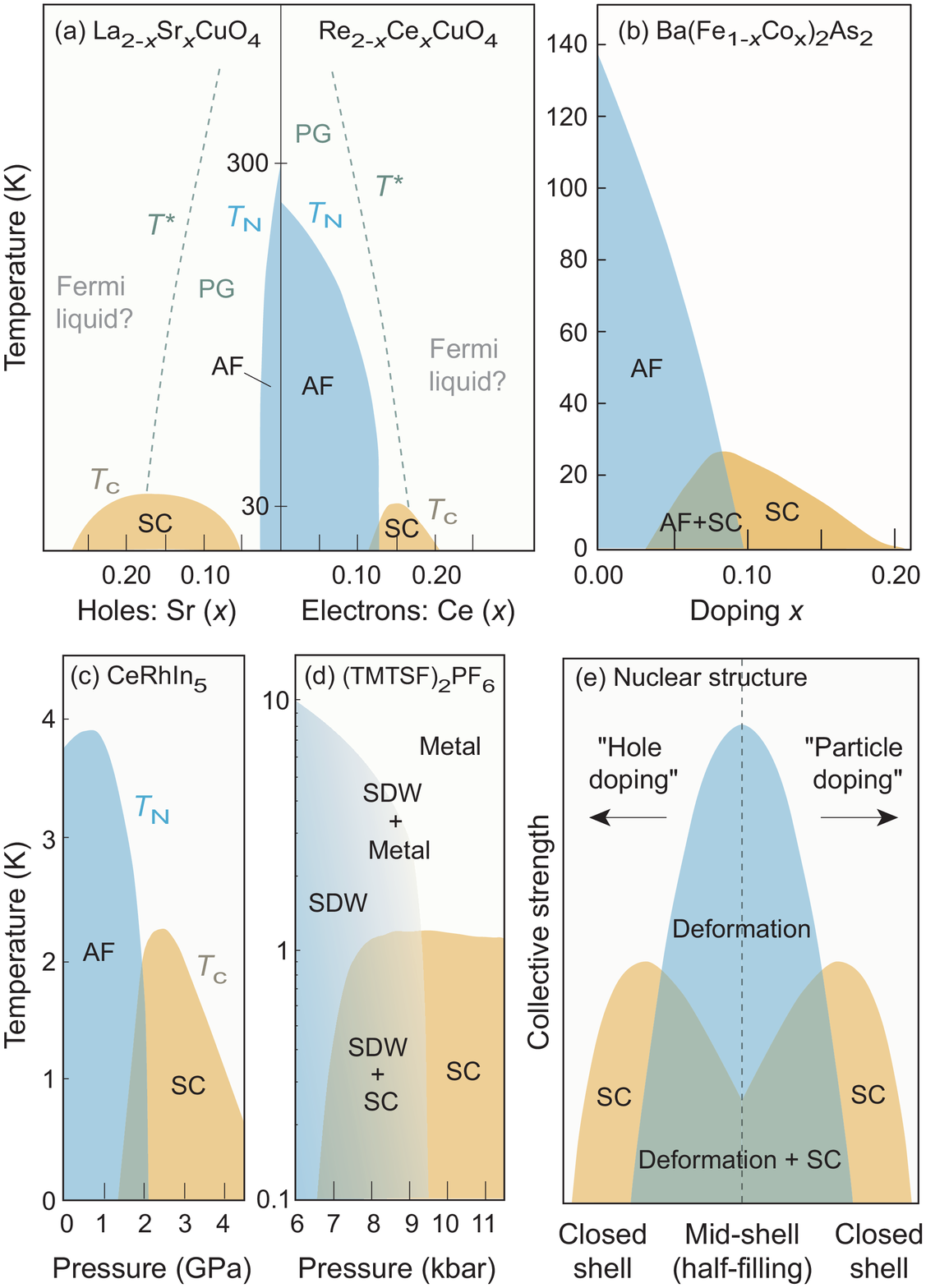}    
{5pt}         
{0pt}         
{0.38}         
{(a)~Phase diagram for  hole- and electron-doped cuprates
\cite{norm2011}. Superconducting (SC), antiferromagnetic (AF), and
pseudogap (PG) regions are labeled, as are  N\'eel ($T\tsub N$), SC
critical ($T\tsub c$), and PG ($T^*$) temperatures. (b)~Phase
diagram for Fe-based SC  \cite{fang09}. (c)~Heavy-fermion phase
diagram \cite{kne09}. (d)~Phase diagram for an organic
superconductor \cite{kang10} (SDW denotes spin density waves).
(e)~Generic correlation-energy diagram for nuclear structure at
$T=0$.}
A striking feature is the proximity of the SC to the
antiferromagnetic (AF) phase. The microscopic pairing structure  is
believed to be dominated by a single band near the Fermi surface and
to have $d_{x^2-y^2}$ orbital geometry.

High-temperature SC in FeAs and FeSe compounds indicates that
cuprate phenomenology like Cu--O planes, $d$-wave pairing, 2D SC,
and Mott-insulator parentage is not essential to high-$T\tsub c$ SC.
A typical phase diagram is shown in \fig{phaseComposite}(b). It is
similar to the cuprate diagram in \fig{phaseComposite}(a), with
adjacent AF and SC phases. The SC and associated pairing in these
systems seems more varied and complex than for the cuprates. For
example, Fe valence-orbital degeneracy suggests that multiple bands
contribute and several orbital geometries may be important for
pairing. Thus, the Fe-based compounds give compelling evidence that
high-$T\tsub c$ SC is compatible with a range of microscopic
structures (a result foreshadowed well before the discovery of
Fe-based SC \cite{guid04}).

A phase diagram for a heavy fermion superconductor is displayed in
\fig{phaseComposite}(c). An AF phase lies adjacent to the SC phase,
as in cuprate and Fe-based phase diagrams. The SC is thought to be
unconventional, and to involve pairs of electrons with effective
masses hundreds of times that of normal electrons.

A phase diagram for an organic superconductor is displayed in
\fig{phaseComposite}(d). It has many similarities with the cuprates,
with pressure replacing doping as the control parameter. The spin
density waves at lower pressure are indicative of AF correlations.
This and many other organic superconductors appear to be
unconventional.

A generic nuclear correlation energy diagram at zero temperature is
shown in \fig{phaseComposite}(e). It is schematic, since  nuclei
have a finite valence space and ``phases'' are mixed by
fluctuations.  Comparing with Figs.\
\ref{fig:phaseComposite}(a)-\ref{fig:phaseComposite}(d) suggests a
strong analogy, with pairing playing a similar role in both cases,
and nuclear quadrupole deformation  being the analog of
condensed-matter antiferromagnetic correlations (an analogy that is
elaborated in \cite{guid01}).

A theory accounting for this diversity of behavior must exhibit
several emergent-state properties: (1)~A robust Cooper instability
arising in both Fermi-liquid and other contexts, depending only
through parameters on microscopic physics. (2)~Accommodation of
SC/SF and other emergent modes, with quantum phase transitions among
these modes. (3)~Limits corresponding to pure SC and to the pure
collective modes that compete with SC. (4)~Limits corresponding to
conventional BCS. (5)~Spontaneous breaking of gauge and possibly
other symmetries in the emergent state.

Unless we assume the great similarity of SC/SF across many disciplines to be 
mere coincidence, the data suggest that superconductivity and superfluidity must 
have a description that can be approximately separated into two parts:  (1)~A 
{\em universal part} describing the essential emergent properties of SC/SF that 
is largely independent of microscopic specifics for the weakly interacting 
parent systems.  (2)~A {\em system-specific part} that can vary from case to 
case and parameterizes the quantitative differences between SC/SF cases, without 
altering substantially the essence of the emergent properties.

The distinction is similar to that between a class and instances of that class 
in object oriented computer programming. The class has a generic description 
specifying the essence of the class that may include parameters having 
unspecified values; various instances of that class then correspond to specific 
implementations (instantiations) of the class with different parameter data 
sets.  Then different instances all inherit the same generic properties of the 
class but may differ from each other quantitatively because they have different 
parameter values.

As a simple example of this concept, consider a class specified by the minimal 
definition of a 2D sphere, with properties corresponding to the radius, 
location, and color defined but with unspecified values. Then multiple instances 
may correspond to spheres having different locations, radii, and colors.  The 
instances differ, yet in a deep sense they are the same, 
since intuitively the specific values of instance parameters for color, radius, 
and location are secondary to the essence of being a sphere.

A theory embodying these features cannot be based directly on microscopic 
properties, since these differ essentially between fields.  The only properties 
that these systems share are that (1)~SC/SF involves Cooper pairs of fermions, 
possibly occurring in the presence of other collective modes,   and (2)~the 
normal system has many degrees of freedom but the SC/SF state is 
phenomenologically simple and so must have only a few effective degrees of 
freedom.

This implies that  the SC/SF state results from an enormous
truncation of the Hilbert space to a simple collective subspace.
The similarity of the SC/SF implies that this subspace is in some
sense the {\em same subspace} across these varied disciplines. The
observed similarities across diverse systems can be ensured if the
collective subspaces corresponding to SC/SF  have the same
symmetries of the Hamiltonian (dynamical symmetries). Then matrix
elements (observables) can be similar across fields because they are
determined by the symmetry, even if the microscopic content of the
wavefunctions and operators (not observables) has little similarity
between disciplines.

Thus, we  propose that all fermionic superconductivity and
superfluidity results from a spontaneous reorganization of the
Hilbert space that transcends microscopic details of the normal
system. The generic structure of this reorganized space accounts for
SC and SF in its myriad forms. Normal-state physics influences the
reorganized space, but only parametrically. The pair condensate
competes for Hilbert space with other emergent modes, suggesting
that conventional BCS states are highly {\em atypical,} representing
limiting cases where the space factorizes and SC/SF decouples from
other emergent modes.

The most powerful means of implementing this space truncation is to
identify dynamical symmetries expressed through Lie algebras and
associated Lie groups
\cite{FDSM,clwu86,guid01,guid04,sun06,guid09,guid10,guid11,IBM,
Bi94,Ia95,wmzha87}. Such methods have been applied extensively in
nuclear \cite{FDSM,IBM}, elementary particle \cite{Bi94}, molecular
\cite{Ia95}, and condensed matter physics \cite{guid01}.  They have
exact many-body solutions for special ratios of coupling strengths,
and approximate solutions  for all coupling strengths using
generalized coherent-state methods.  There is no reason to expect
these dynamical symmetries to be directly related to symmetries of
the weakly-interacting system, since the properties of emergent
collective modes cannot be obtained by power series expansion from
the  parent system.

Such theories  are designed to describe the low-energy collective
states and are likely to fail outside that domain.  Furthermore, on
physical grounds the effective interactions  should vary smoothly
with control parameters such as doping, so rapid local fluctuations
reflecting inadequacies of the dynamical symmetry simplification may
not always be captured.  Thus, such approaches are best suited to
provide simple descriptions of global behavior for highly collective
states.  But that is precisely what is required for our hypothesized
universal part of the SC/SF description.

One might fear that such dynamical symmetry methods imply non-unique candidate
Lie algebras, but their ``quantized'' nature (algebras close only for certain
generator numbers) and generic properties of SC and SF states severely restrict
options. Bound states imply compact groups and the number of low-dimensional
compact Lie groups is small. (Physically, the collective degrees of freedom fit
together consistently only in highly-constrained ways.) Furthermore, a pair
condensate is required, so the algebra must contain both fermion
particle--particle and particle--hole operators, constraining possibilities
further.

A lower limit on generator number follows from counting physical
operators. SC requires spin singlet (or triplet) pairs.  But
collective modes carrying angular momentum (magnetism or quadrupole
fields) mix pairs of different spin.   In condensed matter this
implies  both  singlet and triplet pairs, and a minimum of 8
generators (creation and annihilation operators with spin
degeneracy). In nuclear physics, this corresponds minimally to 12
generators, counting total angular momentum $J=0$ and $J=2$ pairs.
An AF field adds 3 generators, a quadrupole field adds 5, and
conservation of charge and spin or total angular momentum implies 4
additional generators. Adding up, we require minimally  15
generators for condensed matter and 21  for nuclear physics
applications. An upper limit may be estimated by noting that
previous applications to topics as complex as high-$T\tsub c$ SC and
the structure of heavy nuclei required no more than 28 generators
\cite{FDSM,guid01,guid09}.

The {\em only candidate algebras} meeting these conditions with more
than 10 and less than 35 generators are SO(8), Sp(6), SO(7), and
SU(4). The highest symmetries needed for prior nuclear or condensed
matter applications have been SO(8) or Sp(6), with SO(7) and SU(4)
as subalgebras. Thus, we conjecture further that all fermionic SC or
SF derives  from SO(8) or Sp(6) dynamical symmetries. This last
simplification is not essential to our argument, but is consistent
with present knowledge.

We have outlined a universal classification of superconducting and
superfluid behavior, but  we  also require matrix elements for
observables.  Calculation of observables is documented extensively
in the references but we give here representative examples from
condensed matter and nuclear physics.
Figure~\ref{fig:fdsmBE2rareEarth}(a)%
\singlefig {fdsmBE2rareEarth} {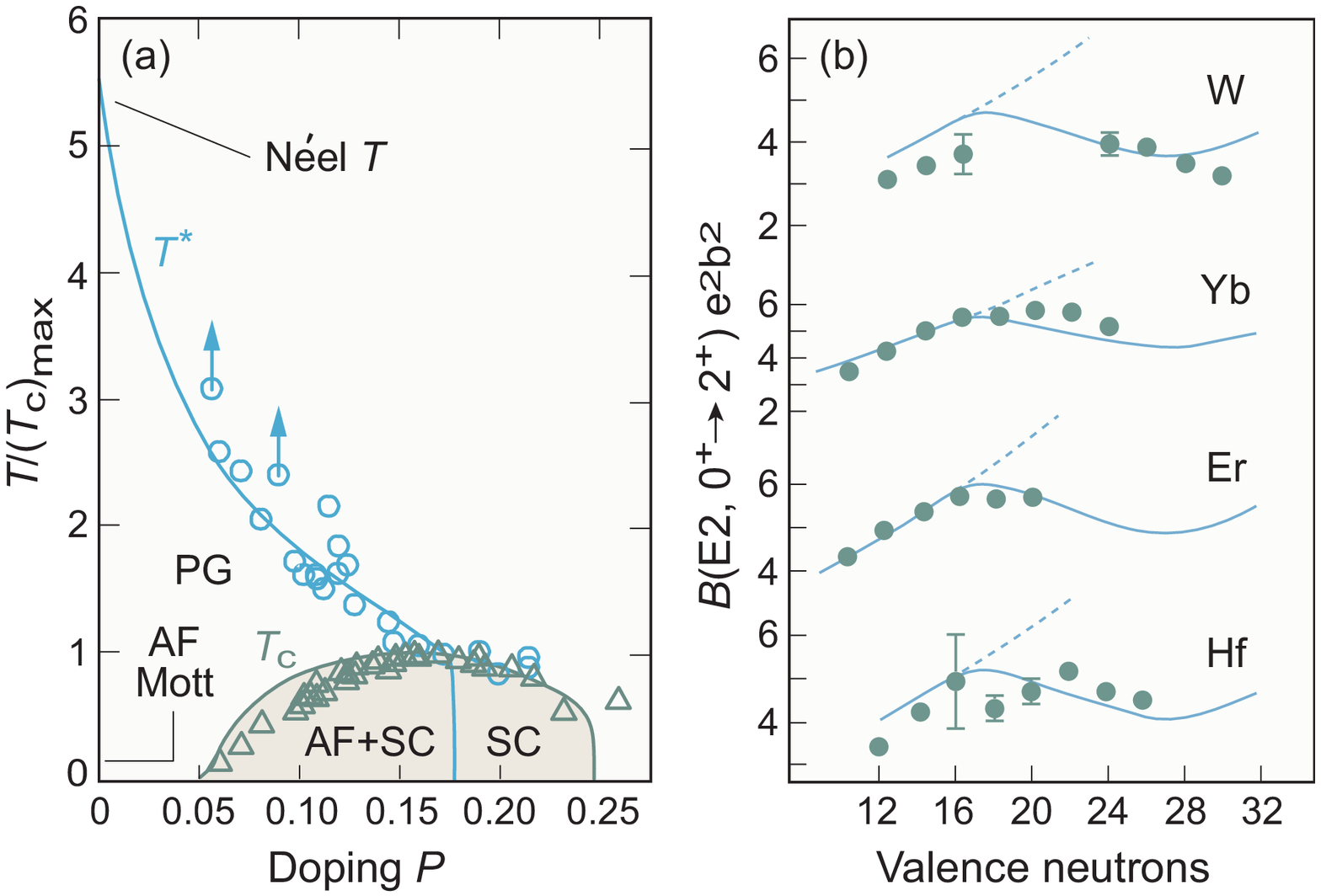} {0pt}
{0pt} {0.44} {(a)~Calculated SU(4) cuprate phase diagram
\cite{sun06}. PG temperature is $T^*$ and SC transition temperature
is $T\tsub c$. Dominant correlations in each region are indicated by
labels SC and AF. Data from  \cite{dai99,camp99}. (b)~Quadrupole
transition rates for rare earth isotopes. Data from \cite{rama87}.
Dashed blue curves correspond to approximating  Cooper pairs as
bosons; clearly they are bosons only for low valence-space
occupancy.}
shows a cuprate phase diagram compared with SU(4)-model calculations
\cite{sun06}. The calculated phase diagram agrees  quantitatively
with data. In Fig.~\ref{fig:fdsmBE2rareEarth}(b) we use the Fermion
Dynamical Symmetry Model (FDSM)  to calculate transition rates
between ground and first excited states in rare earth nuclei
\cite{FDSM}. Again, agreement with data is quite good. Thus, fermion
dynamical symmetries provide both a universal classification  and
methods to calculate observables within specific fields for
superconducting and superfluid behavior, in possible competition
with other collective modes.

Highest symmetries having multiple dynamical symmetry subchains
imply competing ground states and {\em quantum phase transitions}.
The SU(4) model of cuprate SC illustrates.  Because of SU(4)
symmetry, the SC order parameter $\Delta$ satisfies
\begin{equation}
    \left. \frac{\partial\Delta}{\partial x} \right|_{x=0} =
        \left. \frac14 \frac{x_q^{-1} -2x}{(x(x_q^{-1} -x))^{1/2}}
\right|_{x=0}
        = \infty ,
\label{analyticalDelderiv}
\end{equation}
where $x$ is doping and $x_q$ is a critical doping predicted by the
theory: the undoped AF Mott state is unstable against condensing
pairs with infinitesimal doping for finite attractive pairing
\cite{guid10}, as illustrated in \fig{cooperAndAF_Instability}(a).
\singlefig
{cooperAndAF_Instability}
{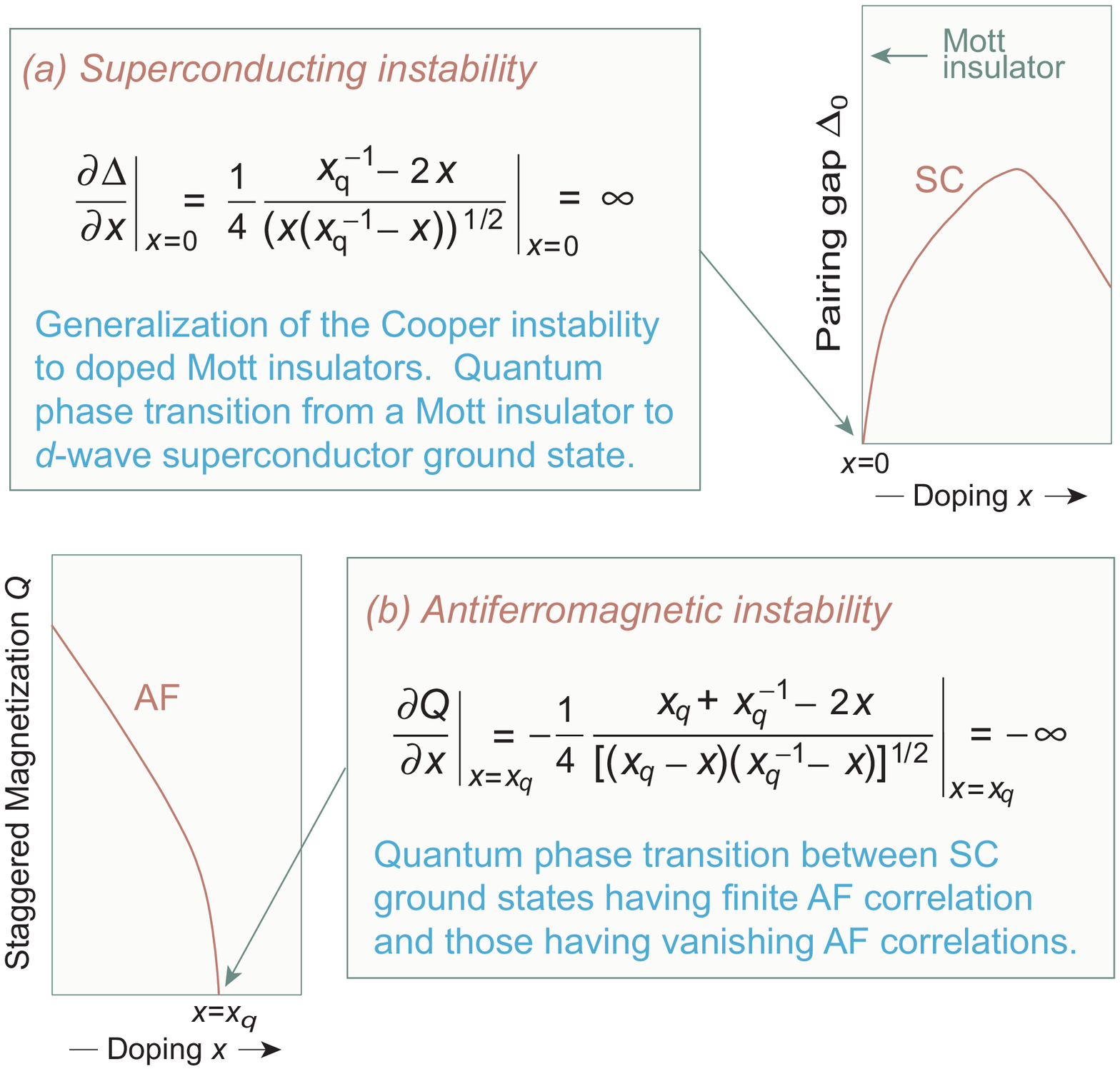}
{5pt}         
{0pt}         
{0.41}         
{ (a)~Cooper instability in a Mott insulator. SU(4)
symmetry requires the ground state at half filling to be an AF Mott
insulator, which for infinitesimal hole doping  becomes unstable against a
quantum phase transition
to a $d$-wave SC state if the  pairing interaction is
finite. (b)~The AF instability of the $d$-wave superconductor.
Because of SU(4) symmetry, as the doping $x$ approaches the critical
doping $x_q$ the system becomes unstable with respect to a quantum
phase transition between an SC state perturbed by AF correlations
for $x<x_q$ and a pure superconductor with no AF correlations for
$x>x_q$. }
The SU(4) symmetry also implies a second fundamental instability:
the AF order parameter $Q$ must satisfy
\begin{equation}
    \left. \frac{\partial Q}{\partial x} \right|_{x=x_q} =
        \left. -\frac14 \frac{x_q + x_q^{-1}-2x}
        {[(x_q-x)(x_q^{-1} -x)]^{1/2}} \right|_{x=x_q}
        = - \infty ,
\label{analyticalDelderiv2}
\end{equation}
and a small change in doping $x$ causes a divergence in AF
correlations if $x \simeq x_q$, as illustrated in
\fig{cooperAndAF_Instability}(b).
In addition, {\em critical dynamical symmetries,} which generalize a
quantum critical point to an entire critical phase and enable a
variety of emergent complexity, have been observed when dynamical
symmetries compete in condensed matter and nuclear systems
\cite{guid11,wmzha87}.

Condensed matter SO(8)
$\supset$ SU(4) and nuclear physics SO(8) and Sp(6) symmetries
reduce to conventional or unconventional BCS SC in the limit where
non-pairing order is neglected \cite{guid04,FDSM}.
\fig{su4-so8_relationship} illustrates for the condensed matter case.%
\singlefig
{su4-so8_relationship}       
{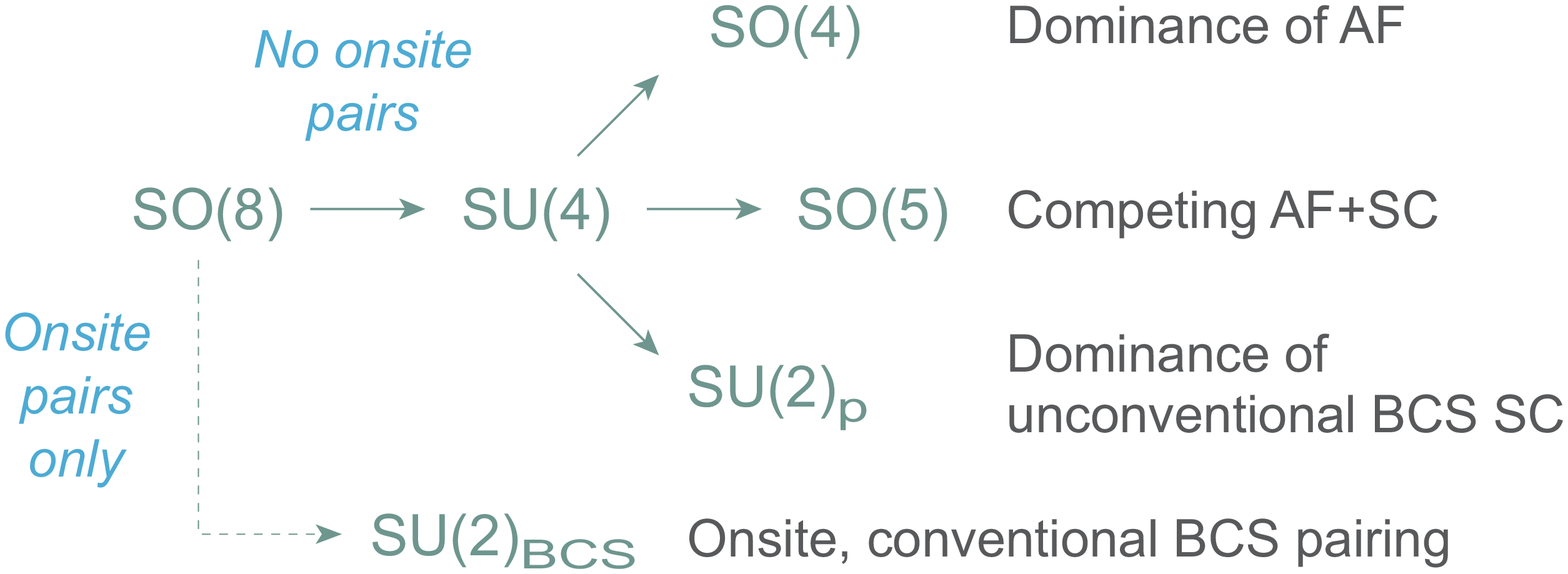}    
{0pt}         
{0pt}         
{0.29}         
{Recovery of BCS states in
a condensed-matter superconductor. Both SU(2)$\tsub{BCS}$ and
SU(2)$\tsub p$ subgroups imply BCS-like states.  They differ in
pairs being onsite for SU(2)$\tsub{BCS}$ and bondwise for
SU(2)$\tsub p$, and in that SU(2)$\tsub{BCS}$ is consistent with
conventional pairing but SU(2)$\tsub p$ can have unconventional
pairing.}
The essential point is not whether SC is conventional or
unconventional, since that influences only the pairing formfactor
and dynamical symmetries are often compatible with a variety of
formfactors \cite{guid04,guid09}.  It is the symmetry of the
truncated Hilbert space that is central to understanding
superconductivity and superfluidity, not  the pairing geometry.

Our proposal has an abstract similarity to general relativity, where
gravity is a universal consequence of spacetime structure, not of
interactions between particles in spacetime. In like manner, the
universality observed for superconductivity and superfluidity across
disciplines derives from the structure of a common Hilbert subspace
selected by  dynamical symmetries.

There also is an analogy with renormalization group flow, since  the dynamical 
symmetries distinguish between ``relevant operators'' characterizing the 
collective subspace and ``irrelevant operators'' that differentiate microscopic 
systems but enter only parametrically into  the collective behavior.  The 
``flow''  is in the dimensionality of the generator space; as it is decreased 
from that of the full Hilbert space toward that of the collective subspace, the 
influence of irrelevant operators falls aside, leaving only relevant operators 
to define the  SC/SF Hilbert subspace. Universality is implied because 
differences between fields are represented by irrelevant operators but the 
relevant operators define SC/SF subspaces having common algebraic structures 
across fields.

Finally, we note that the global view advocated here may illuminate 
specific occurrences of SC and SF in particular subfields.  For example, 
high-$T\tsub c$ cuprate superconductivity becomes far less mysterious when 
viewed as part of a continuum of behavior shared with many other systems.  The 
question of why cuprates differ so much from conventional BCS SC becomes 
inverted in the present view:  it is the {\em conventional BCS superconductors} 
that should more properly be viewed as anomalous, in that they represent only a 
special limit where we may neglect other collective modes competing with SC. 

In summary, a unified understanding of superconductivity and
superfluidity cannot focus on microscopic properties of the normal
state, which are not connected analytically to properties of the
emergent state and may differ radically between disciplines. Nor can
it focus  on Fermi-liquid instabilities, since these phenomena do
not require Fermi-liquid parentage. A common algebraic structure for
the matrix elements  is arguably the only framework that can unify
at the emergent level but depend only parametrically on microscopic
details in such diverse systems. We propose that all fermionic
superconductivity and superfluidity results from a generalized
Cooper instability manifested through fermion dynamical symmetries.
All cases  examined thus far in condensed matter and nuclear physics
derive from  two highest symmetries, SO(8) or Sp(6), suggesting that
an economical unification is possible.

Acknowledgements: We thank Cheng-Li Wu and Lian-Ao Wu for extensive
discussions. This work was partially supported by the Open Project
Program of State Key Laboratory of Theoretical Physics, Institute of
Theoretical Physics, Chinese Academy of Sciences, China.

\end{document}